\documentclass[11pt]{article}
\pdfoutput=1
\usepackage{tikz}
\usetikzlibrary{matrix,arrows,decorations.pathmorphing,decorations.pathreplacing}
\usepackage{jheppub}
\usepackage{amsmath,amssymb,euscript,array,mathrsfs,appendix,ctable,marvosym}
\usepackage{arydshln}
\usepackage{todonotes}
\usepackage{graphicx}
\usepackage[normalem]{ulem}

\input{tcilatex}

\newcommand{\EQ}[1]{\begin{equation}\begin{split} #1
\end{split}\end{equation}}

\title{Holomorphic Chern-Simons theory and lambda models: PCM case}
\author{David M. Schmidtt\footnote{david@df.ufscar.br}} 

\affiliation{Departamento de F\'\i sica, Universidade Federal de S\~ao Carlos, \\
Caixa Postal 676, CEP 13565-905, S\~ao Carlos-SP, Brasil} 

\abstract{In this note we consider the symplectic reduction of a four-dimensional holomorphic Chern-Simons theory recently introduced in \cite{CY} for describing integrable field theories. We work out explicitly the case of the lambda deformed Principal Chiral Model (PCM) and show that the symplectic reduction works as a localization mechanism. The reduced Chern-Simons theory restricts to the set of poles of the twist function underlying the theory, where the known classical integrability of the lambda deformed PCM can be reconstructed from the phase space data associated to this set of points in the spectral space.
\begin{flushleft}
\end{flushleft}
}
\begin{document}

\maketitle


\section{Introduction}

Integrable deformations of string sigma models have attracted a great deal of attention in recent years. Some of the more prominent examples being the Yang-Baxter (or eta) deformations and the lambda deformations, introduced for the PCM in \cite{Klimcik} and\footnote{Although the same action functional appeared earlier in a different context, see \cite{Universal} for the introduction of the action functional and \cite{Thirring} for a proof of its integrability. We thank A. Tseytlin for pointing out these references.} \cite{Sfetsos}, respectively. Both types of deformations were further extended to include other kinds of (super)-string backgrounds and formulations in a series of papers, see for instance \cite{eta-def bos,eta-def fer, eta-def fer 2, Hector,Rene} for the eta deformations and \cite{lambda-bos,lambda-fer,hybrid,PS-lambda} for the lambda deformations. Each deformation have the characteristic of covering a different domain in the deformation parameter space, but are related via Poisson-Lie T-duality and analytic continuation, see \cite{E-models,def-drinfeld}. They are mainly studied for offering a chance to understand the complicated quantum integrable structure of their parent sigma models more efficiently, but as the latter, both types of deformations also belong to the family of the so-called non-ultralocal integrable field theories, were quantization methods like the powerful algebraic Bethe ansatz does not perform well at all. A strategy for eliminating or by-passing this unwanted technical feature, is to embed the theory into a higher dimensional quantum field theory where the non-ultralocality is absent or emerges under some circumstances. In this note, such a higher dimensional field theory will be the four-dimensional Holomorphic Chern-Simons (CS) theory recently introduced in \cite{CY} to formulate integrable field theories\footnote{See \cite{V3}, for a more algebraic approach to handle integrable field theories of the non-ultralocal type.}.

There are at least three major characteristics present in any lambda model suggesting a relation with a gauge theory of the CS type: \\
(i) The presence of two opposite level, mutually commuting Kac-Moody (KM) algebras \cite{Sfetsos,lambda-bos,lambda-fer,hybrid,PS-lambda}, 
\begin{equation}
\big\{\mathscr{L}_{\sigma }(\sigma,z_{\pm} )_{\mathbf{1}},\mathscr{L}_{\sigma }(\sigma^{\prime},z_{\pm} )_{\mathbf{2}}\big\}=\mp
\frac{2\pi }{k}\big( [C_{12},\mathscr{L}_{\sigma }(\sigma^{\prime},z_{\pm} )_{\mathbf{2}%
}]\delta _{\sigma \sigma^{\prime} }+C_{12}\delta^{\prime} _{\sigma \sigma^{\prime} }\big). \label{1 prop}
\end{equation}%
(ii) The factorization (induced by integrability) of the Lagrangian field solution to the equations of motion (eom) in terms of the wave function $\Psi$ \cite{def-drinfeld,lambda background},
\begin{equation}
\mathcal{F} =\Psi (z_{+})\Psi (z_{-})^{-1}. \label{2 prop}
\end{equation}
(iii) The form of the Hamiltonian when expressed in terms of the components of the Lax connection $\mathscr{L}$ \cite{lambdaCS,lambdaCS2},
\begin{equation}
h=\frac{k}{4\pi } \int_{S^{1}}d\sigma \big\langle \mathscr{L}_{\tau }(z_{+})\mathscr{%
L}_{\sigma }(z_{+})-\mathscr{L}_{\tau }(z_{-})\mathscr{L}%
_{\sigma }(z_{-})\big\rangle. \label{3 prop}
\end{equation}
The points $z_{\pm}$ depending on the deformation parameter $\lambda$ are poles of the twist function $\varphi(z)$ of the theory. Indeed, \eqref{1 prop} suggests it in a direct way because of KM algebras rise \cite{Seiberg,zoo}, after symplectic reduction (SR) of a Hamiltonian CS theory defined on a solid cylinder, as Poisson structures of a WZW model defined on its boundary. Equation \eqref{2 prop} mimics the chiral factorization \cite{NA-bos} of the solutions to the eom of an ordinary closed string WZW model and each term in \eqref{3 prop} is identical to the boundary contribution to the canonical Hamiltonian of a CS theory defined on a solid cylinder, if the Lax connection is identified with two of the components of the three-dimensional CS gauge field.

In this note we will focus exclusively on the PCM and consider the problem of how to recover its lambda deformation from the SR of a Hamiltonian CS theory (leaving other models for future work). There are, at least, two possible answers to this question, each one depending fundamentally on the form of the integrand of the symplectic form $\hat{\Omega}$ of the CS theory considered, which is proportional to the two-form
\begin{equation}
\hat{\theta}=\big\langle\hat{\delta}A\wedge\hat{\delta}A \big\rangle, \label{theta}
\end{equation}
with $A$ being the CS gauge field restricted to the constant time manifold $M$ in the decomposition $\mathbb{R}\times M$. The key observation being that the restriction of this two-form to the space of flat connections, taken to be of the form $A=-d\Psi\Psi^{-1}$, is exact
\begin{equation}
\hat{\theta} |_{\text{flat}}=d\big\langle \Psi ^{-1}\hat{\delta} \Psi \wedge d(\Psi
^{-1}\hat{\delta} \Psi )\big\rangle. \label{theta-flat}
\end{equation}
Now we briefly comment on each of the two possibilities for getting a non-trivial reduced symplectic form after integration of the result right above: \\
(I) \textit{``Holography''}. By integrating \eqref{theta-flat} on the disc $M=D$, we obtain the usual result \cite{Seiberg,zoo}
\begin{equation}
\hat{\Omega}_{\text{flat}}\sim \int_{S^{1}}d\sigma \big\langle \Psi ^{-1}\hat{\delta} \Psi \wedge d(\Psi
^{-1}\hat{\delta} \Psi )\big\rangle.
\end{equation}
By considering the addition of two CS actions of opposite levels defined on a solid cylinder, one for each pole $z_{\pm}$, it is possible to recover \eqref{1 prop}, \eqref{2 prop}, \eqref{3 prop} and the lambda deformed PCM action functional as well. This more traditional approach is considered in \cite{lambdaCS, lambdaCS2}, where all the results are presented. The major drawback of this bottom-up approach, is that it is not clear how to include the spectral parameter $z$ in the double CS theory action functional from the very beginning and hence only works partially. \\
In this approach, the SR projects out the degrees of freedom (dof) of the CS theory from the interior of the disc to its boundary inducing some sort of mini-holographic principle and, as a consequence, the reduced theory phase space is determined by the physical data contained on its boundary theory, which turns out to be a lambda model. In this sense, the way the lambda model is recovered is very similar to the way a chiral WZM model is recovered from an usual CS theory. We will not consider this approach in this work. \\
(II) \textit{``Localization''}. By integrating \eqref{theta} on $M=S^{1}\times \mathbb{C}P^{1}$ as follows
\begin{equation}
\hat{\Omega}\sim \int_{S^{1}\times \mathbb{C}P^{1}}\omega\wedge \hat{\theta}, \label{hol CS sf}
\end{equation} 
where $\omega$ is a meromorphic differential defined on $\mathbb{C}P^{1}$, constructed out of the twist function of the underlying integrable field theory, we get something new from \eqref{theta-flat}, i.e. 
\begin{equation}
\hat{\Omega}_{\text{flat}}\sim\int_{S^{1}\times \mathbb{C}P^{1}}d\omega\wedge\big\langle \Psi ^{-1}\hat{\delta} \Psi \wedge d(\Psi \label{flat hol CS}
^{-1}\hat{\delta} \Psi )\big\rangle. 
\end{equation}
The differential $d$ hits $\omega$ and the integral is non-trivial as $d\omega$ is supported at the set of poles $\mathfrak{p}$ of the twist function in the spectral space $\mathbb{C}P^{1}$. The action functional associated to the symplectic form \eqref{hol CS sf} is the four-dimensional holomorphic CS theory first presented in \cite{C1,C2} and subsequently thoroughly studied in a series of papers \cite{W,CWY1,CWY2,CY}. This top-down approach introduce successfully the spectral parameter $z$ into the CS action functional from first principles and everything points towards it is the correct way to do so, as a wide range of known and even new integrable field theories can be described in this way \cite{CY,Vicedo-Unif}, not to mention several lattice integrable models as well. \\
In this approach, and at least for the explicit example to be considered in this note, the SR restricts the degrees of freedom of the CS theory from $M=S^{1}\times \mathbb{C}P^{1}$ to $M=S^{1}\times \mathfrak{p}$ inducing a localization mechanism and, as a consequence, the information of the reduced theory phase space is determined by the restriction of part of the original CS gauge field to the set of poles $\mathfrak{p}$ in the spectral manifold, that is identified with the Lax connection of the lambda model in a natural way. This is the approach that we will consider in what follows.

It is the purpose of this note to work out the approach (II) in detail and to show how \eqref{1 prop}, \eqref{2 prop}, \eqref{3 prop}, the lambda deformed PCM action functional and its classical integrability properties can be recovered from the SR of a holomorphic CS theory defined on $\Sigma\times \mathbb{C}P^{1}$. We emphasize that in this setup the gauge is not fixed completely but only partially, in contrast to \cite{Vicedo-Holo, Vicedo-Unif}, where the CS gauge symmetry is fixed in totality. In this regard our results are complementary. In section \eqref{2} we gather several relevant results of the lambda deformed PCM case and in section \eqref{3} we focus entirely on the approach (II), with the goal of recovering all the results of \eqref{2} from this new perspective. We finish with some comments on the relation between approaches (I) and (II) and under which conditions they describe the same physical system. This is done in section \eqref{4}.

\section{Lambda deformed principal chiral model} \label{2}

In this section we collect some relevant results of the lambda deformed PCM that will facilitate its identification as the reduced field theory obtained by performing a SR on an holomorphic CS theory in the next section. All the results can be found in the literature and are briefly gathered here in order to maintain the text self-contained. The only relatively new detail concerns a differential $\omega$ constructed out of the twist function $\varphi$ of the theory that will play a prominent role in the holomorphic CS theory considered in \eqref{3}.     

\subsection{Action functional and equations of motion}

The lambda deformed PCM is defined by the following action functional\footnote{%
The 1+1 dimensional world-sheet notation used is: $\sigma^{\pm }=\tau\pm \sigma,$ $\partial
_{\pm }=\frac{1}{2}(\partial _{\tau}\pm \partial _{\sigma}),$ $\eta _{\mu \nu
}=diag(1,-1)$, $\epsilon_{01}=1$, $\delta _{\sigma\sigma ^{\prime }}$=$\delta(\sigma-\sigma^{\prime})$, $\delta^{\prime} _{\sigma\sigma ^{\prime }}$=$\partial_{\sigma}\delta(\sigma-\sigma^{\prime})$ and $d^{2}\sigma ^{\prime }\equiv d\sigma ^{-}\wedge d\sigma ^{+}=2d\tau \wedge d\sigma \equiv 2d^{2}\sigma$. Also $a_{\pm }=\frac{1}{2}%
(a_{\tau}\pm a_{\sigma})$ and sometimes we use $\tau=\sigma^{0}$ and $\sigma=\sigma^{1}$ interchangeably.}
\begin{equation}
S_{\lambda}=S_{F/F}(\mathcal{F},A)-\frac{k}{\pi }\dint_{\Sigma
}d^{2}\sigma \big\langle A_{+}(\Omega -1)A_{-}\big\rangle ,
\label{deformed-PCM} 
\end{equation}%
where $\left\langle \ast ,\ast \right\rangle= Tr(\ast ,\ast )$ is the
trace in some faithful representation of the Lie algebra $\mathfrak{f}$, $\Sigma=\mathbb{R}\times S^{1} $ is the closed string world-sheet manifold parameterized by the coordinates $(\tau, \sigma)$, $k$ is the level and 
\begin{equation}
\Omega =\lambda^{-1} I, \text{ \ \ } \lambda^{-1}=1+\frac{\kappa^{2}}{k}
\end{equation}
is the omega projector defining the deformation with $I$ being the identity operator.
Above, we have that%
\begin{equation}
S_{F/F}(\mathcal{F},A)=S_{WZW}(\mathcal{F})_{k}-\frac{k}{\pi }%
\dint_{\Sigma }d^{2}\sigma \big\langle A_{+}\partial _{-}\mathcal{FF}^{-1}-A_{-}%
\mathcal{F}^{-1}\partial _{+}\mathcal{F-}A_{+}\mathcal{F}A_{-}\mathcal{F}%
^{-1}+A_{+}A_{-}\big\rangle ,
\end{equation}%
where $S_{WZW}(\mathcal{F})_{k}$ is the usual level $k$ WZW model action
\begin{equation}
S_{WZW}(\mathcal{F})_{k}=-\frac{k}{2\pi }\int\nolimits_{\Sigma }d^{2}\sigma
\left\langle \mathcal{F}^{-1}\partial _{+}\mathcal{FF}^{-1}\partial _{-}%
\mathcal{F}\right\rangle -\frac{k}{4\pi }\int\nolimits_{\mathcal{M}}\chi(\mathcal{F}^{\prime}) 
\end{equation}%
and
\begin{equation}
\chi(\mathcal{F}^{\prime}) =\frac{1}{3}\big\langle \mathcal{F^{\prime}}^{-1}d\mathcal{F^{\prime}}\wedge \mathcal{F^{\prime}}^{-1}d\mathcal{F^{\prime}}\wedge\mathcal{F^{\prime}}^{-1}d\mathcal{F^{\prime}}\big\rangle
\end{equation}
is the Wess-Zumino three-form defined on a manifold $\mathcal{M}$, where $\Sigma=\partial \mathcal{M}$. The constant $\kappa^2$ is the coupling of the un-deformed PCM.

On the one hand, the $A_{\pm}$ eom are given by\footnote{Set $D=Ad_{\mathcal{F}}$ and $D^{T}=Ad_{\mathcal{F}^{-1}}$.}
\begin{equation}
A_{+}=\left( \Omega ^{T}-D^{T}\right) ^{-1}\mathcal{F}^{-1}\partial _{+}%
\mathcal{F},\text{ \ \ }A_{-}=-\left( \Omega -D\right) ^{-1}\partial _{-}%
\mathcal{FF}^{-1} \label{A in terms of F}
\end{equation}%
and from this follows that the Maurer-Cartan identity for the flat current $\mathcal{F}^{-1}\partial _{\pm}\mathcal{F}$ takes the form
\begin{equation}
\xi_{1}-D^{T}\xi_{2}=0, \label{MC}
\end{equation}
where
\begin{equation}
\xi_{1}=\lbrack \partial _{+}+\Omega^{T}A_{+},\partial _{-}+A_{-}],\text{ \ \ }\xi_{2}=\lbrack \partial _{+}+A_{+},\partial _{-}+\Omega A_{-}]. 
\end{equation}
On the other hand, the $\mathcal{F}$ eom when combined with \eqref{A in terms of F} imply that both terms $\xi_{i}$ in \eqref{MC} vanish separately and 
\begin{equation}
\xi_{1}=\xi_{2}=0. \label{F eom}
\end{equation}

Together, \eqref{A in terms of F} and \eqref{F eom} leads to a system of equations that is formally
equivalent to the PCM eom, i.e.%
\begin{equation}
\partial _{+}I_{-}+\partial _{-}I_{+}=0,\text{ \ \ }\partial
_{+}I_{-}-\partial _{-}I_{+}+[I_{+},I_{-}]=0 \label{pcm eom}
\end{equation}%
but in terms of the deformed dual currents defined by%
\begin{equation}
I_{\pm }=\frac{2}{1+\lambda }A_{\pm }. \label{dual currents}
\end{equation}%
The pair of equations \eqref{pcm eom} follow from the zero curvature condition of the Lax connection%
\begin{equation}
\mathscr{L}_{\pm }(z)=\frac{1}{1\pm z}I_{\pm }, \label{Lax on-shell}
\end{equation}%
or equivalently, as the compatibility of the associated linear problem
\begin{equation}
(\partial_{\mu}+\mathscr{L}_{\mu}(z))\Psi(z)=0,
\end{equation}
where $\Psi$ is the wave function. The latter expression allows to write the Lagrangian fields in the form
\begin{equation}
\begin{aligned}
\mathcal{F} &=\Psi (z_{+})\Psi (z_{-})^{-1},\text{ \ \ \ \ \ \
\ \ \ \ }A_{\pm }=-\partial _{\pm }\Psi (z_{\pm })\Psi (z_{\pm })^{-1}, \\
\Omega ^{T}A_{+} &=-\partial _{+}\Psi (z_{-})\Psi (z_{-})^{-1},\text{ \ \ }%
\Omega A_{-}=-\partial _{-}\Psi (z_{+})\Psi (z_{+})^{-1}, \label{key relations}
\end{aligned} 
\end{equation}
where
\begin{equation}
z_{\pm }=\pm \frac{1-\lambda }{1+\lambda } \label{poles0}
\end{equation}
are two special points in the complex plane (plus a point at infinity) that will play a prominent role in what follows. They are exchanged when we take $\lambda \rightarrow \lambda^{-1}$. From \eqref{Lax on-shell} we notice that the Lax connection vanishes for  $z=\infty$ and \eqref{key relations} is complemented with the wave function boundary condition
\begin{equation}
\underset{z\rightarrow \infty }{\lim }\Psi (z)=Id. \label{psi at infty}
\end{equation}

The equations \eqref{A in terms of F} also take the alternative form
\begin{equation}
\mathscr{J}_{+}=-\frac{k}{2\pi }(\Omega ^{T}A_{+}-A_{-}),\text{ }\mathscr{J}%
_{-}=\frac{k}{2\pi }(A_{+}-\Omega A_{-}), \label{A eom}
\end{equation}%
where we have used the Kac-Moody currents expressions \eqref{KM currents off}. 
In this way, the spatial component of the Lax connection satisfy  
\begin{equation}
\mathscr{L}_{\sigma }(z_{\pm})=\pm \frac{2\pi }{k}\mathscr{J}_{\mp },
\end{equation}%
where the $\mathscr{J}_{\pm }$ are to be taken as the right hand sides in both equations of \eqref{A eom}. 

Using \eqref{A eom} and \eqref{dual currents}, we obtain expressions \cite{quantum-group} for the space and time components of the
Lax connection in terms of the Kac-Moody currents. Namely,
\begin{equation}
\mathscr{L}_{\sigma }(z)=f_{+}(z)\mathscr{J}_{+}+f_{-}(z)\mathscr{J}_{-}, \text{ \ \ }\mathscr{L}_{\tau }(z)=g_{+}(z)\mathscr{J}_{+}+g_{-}(z)\mathscr{J}_{-}, \label{spatial-time Lax}
\end{equation}%
where we have defined the functions%
\begin{equation}
f_{\pm }(z)=\alpha \frac{(z-z_{\pm })}{(1-z^{2})},\text{ \ \ }g_{\pm }(z)=\alpha z_{\pm } \frac{(z-z_{\pm }^{-1})}{(1-z^{2})} \label{efes y ges}
\end{equation}%
and the constant
\begin{equation}
\alpha =\frac{4\pi }{k}\frac{\lambda}{1-\lambda^{2} }.
\end{equation}
Equation \eqref{spatial-time Lax} is important for understanding the integrable properties of the lambda model as well as its relation to the holomorphic CS theory. Indeed, as follows from \eqref{spatial-time Lax}, the quantity
\begin{equation}
B(z)=\epsilon ^{\mu \nu }\big\langle \mathscr{L}_{\mu }(z)\delta \mathscr{L}_{\nu }(z)\big\rangle\sim \frac{1}{z^{2}-1} \label{bdry z}
\end{equation}%
satisfy
\begin{equation}
B(z_{+})-B(z_{-})=0,\text{ \ \ }\underset{z\rightarrow \infty }{\lim }\partial_{z}B(z)=0. \label{bdry eom}
\end{equation}%
This condition imply that the variational problem in the holomorphic CS theory is well-defined. In \eqref{bdry z} the symbol $\sim$ denotes the $z$ dependence.

Finally, in order to see if the Lax connection \eqref{Lax on-shell} is flat, not only on-shell but off-shell as well, we must run the Dirac algorithm first. This we do next.

\subsection{Hamiltonian structure and integrability}

The phase space associated to the action functional \eqref{deformed-PCM} is described by the following data: two currents $\mathscr{J}_{\pm }$ given by
\begin{equation}
\mathscr{J}_{+}=-\frac{k}{2\pi }\left( \mathcal{F}^{-1}\partial _{+}\mathcal{%
F+F}^{-1}A_{+}\mathcal{F-}A_{-}\right) ,\text{ \ \ }\mathscr{J}_{-}=\frac{k}{%
2\pi }\left( \partial _{-}\mathcal{F\mathcal{F}}^{-1}\mathcal{-F}A_{-}%
\mathcal{F}^{-1}\mathcal{+}A_{+}\right)  \label{KM currents off}
\end{equation}%
that obey the relations of two opposite levels mutually commuting Kac-Moody algebras\footnote{For the Lie algebra we use the definitions:
$\eta _{AB}=\left\langle T_{A},T_{B}\right\rangle ,$ $%
C_{\mathbf{12}}$ = $\eta ^{AB}T_{A}\otimes T_{B}$ and $u_{\mathbf{1}}=u\otimes I,$ $u_{\mathbf{2}}%
=I\otimes u$.}
\begin{equation}
\big\{\mathscr{J}_{\pm }(\sigma)_{\mathbf{1}},\mathscr{J}_{\pm }(\sigma^{\prime})_{\mathbf{2}}\big\}=%
-[C_{\mathbf{12}},\mathscr{J}%
_{\pm }(\sigma^{\prime})_{\mathbf{2}}]\delta _{\sigma\sigma^{\prime}}\pm \frac{k}{2\pi }C_{\mathbf{12}}\delta _{\sigma\sigma^{\prime}}^{\prime } \label{KM algebra}
\end{equation}%
and two conjugated  pairs of fields $(A_{\pm},P_{\mp})$ with Poisson brackets
\begin{equation}
\big\{P_{\pm }(\sigma)_{\mathbf{1}},A_{\mp }(\sigma^{\prime})_{\mathbf{2}}\big\}=\frac{1}{2}C_{\mathbf{12}}\delta _{\sigma\sigma^{\prime}}.
\end{equation}
The time evolution is determined by the canonical Hamiltonian density
\begin{equation}
H_{C}=-\frac{k}{\pi }\Big\langle \left( \frac{\pi }{k}\right) ^{2}\left( 
\mathscr{J}_{+}^{2}+\mathscr{J}_{-}^{2}\right) -\frac{2\pi }{k}\left( A_{+}%
\mathscr{J}_{-}+A_{-}\mathscr{J}_{+}\right) +\frac{1}{2}\left(
A_{+}^{2}+A_{-}^{2}\right) -A_{+}\Omega A_{-}\Big\rangle \label{canonical H}
\end{equation}%
through the relation 
\begin{equation}
\partial _{\tau }f =\big\{h_{C},f\big\} ,%
\text{ \ \ }h_{C}=\int\nolimits_{S^{1}}d\sigma H_{C}(\sigma
), \label{flow in lambda}
\end{equation}
where $f$ is an arbitrary functional of the phase space variables.

Now we consider the Dirac algorithm. There are two primary constraints
\begin{equation}
P_{+}\approx 0,\text{ \ \ }P_{-}\approx 0. 
\end{equation}%
By adding them to the canonical Hamiltonian density we construct the total Hamiltonian density
\begin{equation}
H_{T}=H_{C}-2\big\langle u_{+}P_{-}+u_{-}P_{+}\big\rangle,
\end{equation}%
where $u_{\pm}$ are arbitrary Lagrange multipliers. 

The time preservation of the primary constraints under the flow of $H_{T}$ produces two secondary constraints given by
\begin{equation}
C_{+}=\mathscr{J}_{+}+\frac{k}{2\pi }\left( \Omega ^{T}A_{+}-A_{-}\right)
\approx 0,\text{ \ \ }C_{-}=\mathscr{J}_{-}-\frac{k}{2\pi }\left(A_{+}-\Omega
A_{-}\right) \approx 0,  \label{secondary const}
\end{equation}%
which are the gauge field eom \eqref{A eom} found above.
By adding these secondary constraints to the total Hamiltonian we construct the extended Hamiltonian
\begin{equation}
H_{E}=H_{C}-2\left\langle u_{+}P_{-}+u_{-}P_{+}+\mu_{+}C_{-}+\mu_{-}C_{+}\right\rangle,
\end{equation}
where $\mu_{\pm}$ are arbitrary Lagrange multipliers. 

Verifying again the preservation of the primary and secondary constraints under the flow of $H_{E}$, leads to the complete determination of the Lagrange multipliers and no new tertiary constraints produced at this level. 

Before we consider the Virasoro constraints, it is useful to separate the constraints we have found so far between first and second class constraints in order to simplify the rest of the analysis. We quickly realize that there are no first class constraints because of the pairs
\begin{equation}
P_{\pm }\approx 0\text{ \ \ and \ \ }C_{\pm }\approx 0 \label{second class pair}
\end{equation}
form a system of second class pairs of constraints. We impose them strongly by
means of a Dirac bracket. However, the Poisson brackets among the currents $\mathscr{J}_{\pm}$ are not modified \cite{lambda-bos}, so we continue using their usual Kac-Moody Poisson brackets \eqref{KM algebra}. As a consequence, the expressions \eqref{A eom} and \eqref{spatial-time Lax} are valid in the strong sense. At this point, we can anticipate that no Hamiltonian extension of the Lax connection will be required in contrast to the lambda models on (semi)-symmetric spaces \cite{lambdaCS2}.

Now, we are ready to consider the Virasoro constraints which must be imposed by hand in the conformal gauge approach adopted here. After a temporary reintroduction of the 2d world-sheet metric in the action \eqref{deformed-PCM}, we find the stress-tensor components
\begin{equation}
T_{\pm \pm }=-\frac{k}{4\pi }\big\langle \left( \mathcal{F}^{-1}D_{\pm }%
\mathcal{F}\right) ^{2}+2A_{\pm }(\Omega -1)A_{\pm }\big\rangle ,
\end{equation}%
where $D_{\pm }(\ast )=\partial _{\pm }(\ast )+\left[ A_{\pm },\ast \right]$ is a covariant derivative and
after imposing \eqref{secondary const} strongly, we find that
\begin{equation}
T_{\pm \pm }=-\frac{k}{16\pi }\frac{(1+\lambda )^{3}(1-\lambda )}{\lambda
^{2}}\left\langle I_{\pm }I_{\pm }\right\rangle. \label{Virasoro}
\end{equation}
Above, the currents $I_{\pm}$ are given by \eqref{dual currents}. Another expression is given in terms of the Lax connection \eqref{spatial-time Lax} and the points \eqref{poles0}, i.e.
\begin{equation}
T_{\pm \pm }=\pm \frac{k}{4\pi }\left\langle \mathscr{L}_{\pm }^{2}(z_{+})-%
\mathscr{L}_{\pm }^{2}(z_{-})\right\rangle .
\end{equation}
From these results, it is straightforward to show that
\begin{equation}
\begin{aligned}
\big\{p_{+}, \mathscr{L}_{-}(z)\big\}-\big\{ p_{-},
\mathscr{L}_{+}(z)
\big\}=-\big[\mathscr{L}_{+}(z),\mathscr{L}_{-}(z)\big],
\end{aligned}
\end{equation}%
where
\begin{equation}
p_{\pm }=\int\nolimits_{S^{1}}d\sigma T_{\pm \pm
}(\sigma ),
\end{equation}
confirming that the Lax pair $\mathscr{L}_{\pm }(z)$ is a strongly flat $z$-dependent connection and that no Hamiltonian extension is required, i.e. the equation
\begin{equation}
\partial _{+}\mathscr{L}_{-}(z)-\partial _{-}\mathscr{L}%
_{+}(z)+\big[ \mathscr{L}_{+}(z),\mathscr{L}_{-}(z)%
\big] =0 \label{strong flatness}
\end{equation}
holds on the whole phase space of this lambda model. 

Because of the Lax connection is strongly flat, the relations \eqref{key relations} are valid off-shell and in terms of the variables $\Psi(z_{\pm})$, the action \eqref{deformed-PCM} takes the form \cite{lambdaCS2}
\begin{equation}
S_{\text{eff} }=-\frac{k}{4\pi }\dint\nolimits_{\mathcal{M}}[\chi (\Psi^{\prime} (z_{+}))-\chi
(\Psi^{\prime} (z_{-}))]. \label{lambda model action}
\end{equation}%
As the constraints $C_{\pm}\approx 0$ have been imposed strongly, the action right above is equivalent to the effective action of the lambda model in the deformed metric and antisymmetric field (i.e. with the field $A_{\pm}$ in \eqref{deformed-PCM} integrated out). In this guise, the action \eqref{lambda model action} is manifestly invariant under the symmetry $\lambda\rightarrow \lambda^{-1}$ and $k\rightarrow-k$ or, equivalently,
\begin{equation}
z_{\pm}\rightarrow z_{\mp},\text{ \ \ }k\rightarrow-k. \label{rare symmetry}
\end{equation} 
We will return to this symmetry later on. 

The remaining constraints left are (the first class) Virasoro's $T_{\pm \pm}\approx 0$, whose action on the transport matrix
\begin{equation}
T( \sigma _{2},\sigma _{1}) =P\exp \big[ -\int\nolimits_{\sigma
_{1}}^{\sigma _{2}}d\sigma \mathscr{L}_{\sigma }(\sigma ;z)\big] \label{transport}
\end{equation}%
is given
\begin{equation}
\left\{p_{\pm}, T(\sigma _{2},\sigma _{1})\right\} =T(\sigma _{2},\sigma _{1})\mathscr{L}_{\pm }(\sigma_{1};z)-\mathscr{L}%
_{\pm }(\sigma_{2};z)T(\sigma _{2},\sigma _{1}).
\end{equation}
The Hamiltonian and momentum densities are
\begin{equation}
H=H_{C}=T_{++}+T_{--}, \text{ \ \ } P=T_{++}-T_{--}
\end{equation}
and from this follows that the trace of powers of the monodromy matrix
\begin{equation}
m(z)=P\exp \big[ -\oint\nolimits_{S^{1}}d\sigma \mathscr{L}%
_{\sigma }(\sigma ;z)\big] , \label{monodromy}
\end{equation}%
is conserved in time. In terms of the Lax connection, the Hamiltonian and momentum densities take the form
\EQ{
H &=\frac{k}{4\pi }\big\langle \mathscr{L}_{\tau }(z_{+})\mathscr{%
L}_{\sigma }(z_{+})-\mathscr{L}_{\tau }(z_{-})\mathscr{L}%
_{\sigma }(z_{-})\big\rangle , \\ 
P &=\frac{k}{8\pi }\big\langle (\mathscr{L}_{\tau }^{2}(z_{+})+%
\mathscr{L}_{\sigma }^{2}(z_{+}))-(\mathscr{L}_{\tau }^{2}(z_{-})+\mathscr{L}_{\sigma }^{2}(z_{-}))\big\rangle . \label{densities}
}

As a consequence of the KM algebra structure, the classical exchange algebra of the theory takes the Maillet's algebra form \cite{Maillet}%
\begin{equation}
\begin{aligned}
\big\{\mathscr{L}_{\sigma}(\sigma;z)_{\mathbf{1}},\mathscr{L}_{\sigma}(\sigma^{\prime};z^{\prime})_{\mathbf{2}}\big\}=[\mathfrak{r}&
_{\mathbf{12}}(z,z^{\prime}),\mathscr{L}_{\sigma}(\sigma;z)_{\mathbf{1}}+\mathscr{L}_{\sigma}(\sigma^{\prime};z^{\prime})_{\mathbf{2}}]\delta
_{\sigma \sigma^{\prime}} \\
+&[\mathfrak{s}_{\mathbf{12}}(z,z^{\prime}),\mathscr{L}_{\sigma}(\sigma;z)_{\mathbf{1}}-\mathscr{L}_{\sigma}(\sigma^{\prime};z^{\prime})_{\mathbf{2}}]\delta _{\sigma \sigma^{\prime}}-2\mathfrak{s}_{\mathbf{12}}(z,z^{\prime})\delta _{\sigma \sigma^{\prime}}^{\prime }, \label{MailletPCM}
\end{aligned}
\end{equation}%
where%
\begin{equation}
\mathfrak{r}_{\mathbf{12}}(z,z^{\prime})=-\frac{\left[ \varphi
^{-1}(z^{\prime})+\varphi^{-1}(z)\right] }{z-z^{\prime}}C_{\mathbf{12}},\text{ \
\ }\mathfrak{s}_{\mathbf{12}}(z,z^{\prime})=-\frac{\left[ \varphi^{-1}(z^{\prime})-\varphi ^{-1}(z)\right] }{z-z^{\prime}}C_{\mathbf{12}} \label{r and s}
\end{equation}%
and $\varphi^{-1}(z)$ is the inverse of the twist function of the model
\begin{equation}
\varphi(z)=\frac{2}{\alpha}\frac{(1-z^{2})}{(z^{2}-z_{+}^{2})}. \label{1 twisting}
\end{equation}%

Now, in order to make the connection with the holomorphic CS theory below more transparent, we take a closer look to the twist function. 

Define the $1$-form
\begin{equation}
\omega=\varphi(z)dz \label{1-form}
\end{equation}
and introduce a $\mathbb{C}P^{1}$ spectral space\footnote{The 2 dimensional complex notation used is: $z=z^{0}+iz^{1},$ $\overline{z}%
=z^{0}-iz^{1},$ $\partial _{z}=\frac{1}{2}\left( \partial _{0}-i\partial
_{1}\right) ,$ $\partial _{\overline{z}}=\frac{1}{2}\left( \partial
_{0}+i\partial _{1}\right) ,$ $\eta _{ab}=diag(1,1),$ $\epsilon _{01}=1,$
$\eta _{z\overline{z}}=\frac{1}{2},$ $\epsilon _{z\overline{z}}=\frac{i}{2}$, $\delta_{zz^{\prime}}=\delta(z-z^{\prime})$
and $dz\wedge d\overline{z}=-2idz^{0}\wedge dz^{1}$.} parameterized by the holomorphic coordinate $z$, with $z$ being the spectral parameter of theory.
A very important result involve the differential two-form $d\omega$ and its corresponding support at the set of poles of the twist function given by 
\begin{equation}
\mathfrak{p}=\{z_{+},z_{-},\infty\}. \label{poles}
\end{equation}
To see this explicitly, expand \eqref{1-form} locally around the points in $\mathfrak{p}$ and keep only the singular contributions. We get\footnote{In each term, the coordinate $z$ is to be understood as a local coordinate around the corresponding pole.}
\begin{equation}
\omega =\frac{k}{\pi }\frac{dz}{z-z_{+}}-\frac{k}{\pi }\frac{dz}{z-z_{-}}+\omega_{\infty}
\end{equation}%
and from this follows that%
\begin{equation}
d\omega =-2kidz\wedge d\overline{z}\left\{ \delta _{zz_{+}}-\delta
_{zz_{-}}\right\}+d\omega_{\infty} ,
\end{equation}%
where%
\begin{equation}
\delta _{zz^{\prime}}=\frac{1}{2\pi i}\frac{\partial }{\partial \overline{z}}\left( 
\frac{1}{z-z^{\prime}}\right) 
\end{equation}%
is the Dirac delta function with the property that
\begin{equation}
\int\nolimits_{\mathbb{C}P^{1}}dz\wedge d\overline{z}F(z)\delta _{zz^{\prime}}=F(z^{\prime}),
\end{equation}%
for any $F\in C^{\infty }(%
\mathbb{C}
P^{1})$. In this way, we get an useful formula\footnote{We have discarded the contribution at $\infty$ because any $F$ is constructed out of the components of the Lax connection, which vanish at that point.}
\begin{equation}
\int\nolimits_{%
\mathbb{C}
P^{1}}d\omega \text{ }F(z)=-2ki\{F(z_{+})-F(z_{-})\}=-2\pi i\sum\limits_{x\in \mathfrak{p}%
}res_{x}\omega F, \label{useful}
\end{equation}
where we have used the definition
\begin{equation}
res_{z_{\pm }}\omega F=\underset{z\rightarrow z_{\pm }}{\lim }%
(z-z_{\pm })\varphi (z)F(z)=\pm \frac{k}{\pi }F(z_{\pm }). \label{residues}
\end{equation}

Armed with these results, we write the Hamiltonian and momentum functions \eqref{densities} in the form
\begin{equation}
h =\frac{1}{4}\sum_{x\in \mathfrak{p}}res_{x}\omega \int\nolimits_{S^{1}}d\sigma \big\langle \mathscr{L}%
_{\tau }\mathscr{L}_{\sigma }\big\rangle  ,\text{ \ \ }
p =\frac{1}{8}\sum_{x\in \mathfrak{p}}res_{x}\omega \int\nolimits_{S^{1}}d\sigma \big\langle \mathscr{L}%
_{\tau }^{2}+\mathscr{L}_{\sigma }^{2}\big\rangle . \label{H and P as res}
\end{equation}%
Below we will show, that the time evolution in the symplectic reduced CS field theory is dictated by $h$. There, a clear interpretation of the expression \eqref{lambda model action} will be given as well.

\section{Holomorphic Chern-Simons theory}\label{3}

In this section we recover the results of \eqref{2} from the holomorphic CS theory point of view. It is important to emphasized that this is done without fixing the gauge symmetry of the CS theory completely (as in \cite{Vicedo-Holo,Vicedo-Unif}), but rather from a symplectic reduction perspective (as in \cite{lambdaCS,lambdaCS2,Aubin}). The main result is that in the holomorphic CS theory case, the SR works as a localization mechanism that eliminates the spectral parameter from the reduced CS theory phase space, which is identified as being equivalent to the lambda model. As a consequence, important quantities of the lambda deformed PCM, like the Lax connection, action functional, exchange algebra and so on, are determined by the phase space data associated to the set of poles where the theory localize in the $\mathbb{C}P^{1}$ spectral space.     

\subsection{Action functional and equations of motion}

The holomorphic Chern-Simons theory of our interest is defined by the following four-dimensional action functional\footnote{The normalization here is determined by the twist function \eqref{1 twisting} and by the condition of recovering \eqref{lambda model action} after performing the SR.}
\begin{equation}
S_{CS}=\frac{i}{8\pi }\int\nolimits_{\Sigma \times \mathbb{C}P^{1}}\omega \wedge CS(B),\text{\ \ \ } CS(B)=\big\langle B\wedge \hat{d}B+\frac{2}{3}B\wedge B\wedge B\big\rangle,\label{Hol CS action}
\end{equation} 
where $\Sigma=\mathbb{R}\times S^{1}$ is the closed string world-sheet manifold, $\omega$ is as defined in \eqref{1-form}, $CS(B)$ is the CS three-form for the gauge field $B$ and $\mathbb{C}P^{1}$ is the spectral space introduced above. Under certain circumstances, as considered in \cite{Vicedo-Unif}, the action is real.

The gauge field and the exterior derivative decompose in the form
\begin{equation}
\begin{aligned}
B &=A_{\tau }d\tau +A,\text{ \ \ \: }A=A_{\sigma }d\sigma +A_{\overline{z}}d%
\overline{z}, \\
\hat{d} &=d\tau \wedge \partial _{\tau }+d,\text{ \ \ }d=d\sigma \wedge
\partial _{\sigma }+dz\wedge \partial _{z}+d\overline{z}\wedge \partial _{%
\overline{z}},
\end{aligned} \label{quantities}
\end{equation}%
where we have ignored the $A_{z}dz$ component of the gauge field $A$ as it completely decouples from the theory. This is because of the 1-form $\omega$ already carries the $dz$ factor contribution to the volume form of $\Sigma \times \mathbb{C}P^{1}$. 

Under the gauge symmetry transformations%
\begin{equation}
B_{g}=gBg^{-1}-\hat{d}gg^{-1}, \label{gauge transf}
\end{equation}%
the CS three-form changes as follows%
\begin{equation}
CS(B_{g})=CS(B)+\chi(g)+\hat{d}\big\langle g^{-1}\hat{d}g\wedge B\big\rangle \label{CS under gauge}
\end{equation}
and, in principle, the theory \eqref{Hol CS action} will be gauge invariant provided the following two conditions are satisfied,
\begin{equation}
\frac{i}{8\pi }\int\nolimits_{\Sigma \times 
\mathbb{C}
P^{1}}\omega \wedge \chi (g)=2\pi N\text{ \ \ and \ \ }g|_{\mathfrak{p}}=Id. \label{gauge conditions}
\end{equation}
We will analyze these conditions more closely from another perspective below.

In the variables \eqref{quantities}, the action becomes
\begin{equation}
S_{CS}=\frac{i}{8\pi }\int\nolimits_{\Sigma \times 
\mathbb{C}
P^{1}}d\tau \wedge \omega \wedge \big\langle A\wedge \partial _{\tau
}A-2A_{\tau }F\big\rangle +\frac{i}{8\pi }\int\nolimits_{\Sigma \times 
\mathbb{C}
P^{1}}d\tau \wedge d\omega \wedge \big\langle A_{\tau }A\big\rangle . \label{Hol-CS-decomp}
\end{equation}%
The Lagrangian of the theory is given by 
\begin{equation}
L=\frac{i}{8\pi }\int\nolimits_{S^{1} \times 
\mathbb{C}
P^{1}}\omega \wedge \big\langle A\wedge \partial _{\tau
}A-2A_{\tau }F\big\rangle +\frac{i}{8\pi }\int\nolimits_{S^{1} \times 
\mathbb{C}
P^{1}}d\omega \wedge \big\langle A_{\tau }A\big\rangle 
\end{equation}%
and has an arbitrary variation of the form
\begin{equation}
\begin{aligned}
\delta L= \frac{i}{4\pi }\int\nolimits_{S^{1} \times \mathbb{C}P^{1}} \omega \wedge \big\langle \delta A \wedge (\partial_{\tau }A-DA_{\tau })-\delta A_{\tau }F\big\rangle +\frac{i}{8\pi }%
\int\nolimits_{S^{1} \times 
\mathbb{C}
P^{1}}d\omega \wedge \big\langle A\delta A_{\tau }-A_{\tau
}\delta A\big\rangle , \label{variation}
\end{aligned}
\end{equation}%
where $D(\ast )=d(\ast )+[A,\ast ]$ is a covariant derivative and where $F=dA+A\wedge A$ is the
field strength for the gauge field $A$.

The eom of the theory follow directly from \eqref{variation}. The $z$-dependent ``bulk" eom given by\footnote{Actually, these eom are to be wedged with $\omega$ but at this point it is already understood the connection is flat but varying holomorphically in $\mathbb{C}P^{1}$.}
\begin{equation}
F=0,\text{ \ \ }\partial _{\tau }A-DA_{\tau }=0 \label{bulk eom}
\end{equation}%
must be supplemented with the ``boundary" condition (cf. footnote 1)
\begin{equation}
\sum_{x\in \mathfrak{p}}\epsilon ^{\mu \nu }res_{x}\omega \big\langle
A_{\mu }\delta A_{\nu }\big\rangle =0 \text{ \ \ for \ \ } \mu=\tau,\sigma. \label{bdry CS}
\end{equation}%
We will refer to these kind of expressions as boundary contributions \cite{CY}. The \eqref{bdry CS} is identical to the true geometrical boundary contribution that appear in the double CS theory approach to lambda models of \cite{lambdaCS,lambdaCS2}, so this name is appropriate in both approaches. In contrast, we will refer to the other type of contributions simply as bulk contributions.

The condition \eqref{bdry CS} is equivalent to \eqref{bdry eom} provided we make the identifications%
\begin{equation}
A_{\mu }(z_{\pm })=\mathscr{L}_{\mu }(z_{\pm }). \label{A=L at poles}
\end{equation}%
In what follows, we will assume this boundary condition is always satisfied and furthermore, we will extend \eqref{A=L at poles} to be valid not only at the points $z_{\pm}$ but at any other value of $z$ as well, i.e.
\begin{equation}
A_{\mu }(z)=\mathscr{L}_{\mu }(z). \label{A=L}
\end{equation}
The proper justification of the key relation \eqref{A=L} requires the use of the Hamiltonian analysis, which is our next topic.

\subsection{Hamiltonian structure and symplectic reduction}

The phase space associated to the Lagrangian \eqref{Hol-CS-decomp} is described by the following
data: three conjugate pairs of fields $(A_{i}, P_{i})$, $i =\tau, \sigma, \overline{z} $ obeying the fundamental
Poisson bracket relations
\begin{equation}
\big\{ A_{i}(\sigma ,z)_{\mathbf{1}},P_{i}(\sigma^{\prime}
,z^{\prime}
)_{\mathbf{2}}\big\} =C_{\mathbf{12}}\delta _{\sigma \sigma^{\prime}}\delta _{zz^{\prime}} \label{PB CS}
\end{equation}%
and a time evolution determined by the canonical Hamiltonian%
\begin{equation}
h_{C}=\frac{i}{4\pi }\dint\nolimits_{S^{1}\times \mathbb{C} P^{1}}\omega \wedge \big\langle A_{\tau }F\big\rangle -\frac{i}{8\pi }%
\dint\nolimits_{S^{1}\times \mathbb{C} P^{1}}d\omega \wedge \big\langle A_{\tau }A\big\rangle , \label{can Ham CS}
\end{equation}%
through the relation 
\begin{equation}
\partial _{\tau }f=\left\{ f,h_{C}\right\} , \label{flow in CS}
\end{equation}%
where $f$ is an arbitrary function of the phase space variables.

Because of the condition \eqref{bdry CS} is assumed to apply, the canonical Hamiltonian has a well-defined functional variation, in the sense that no boundary contributions appear \cite{Regge-Teitelboim}, i.e.
\begin{equation}
\delta h_{C}=\frac{i}{4\pi }\dint\nolimits_{S^{1}\times \mathbb{C} P^{1}}\omega \wedge \big\langle \delta A_{\tau }F+\delta A \wedge D A_{\tau}\big\rangle,
\end{equation}%
or more explicitly,
\begin{equation}
\delta h_{C}=\frac{i}{4\pi }\int_{S^{1}\times 
\mathbb{C}
P^{1}}d_{Vol}\big\langle \delta A_{\tau }\left( \varphi F_{\overline{z}%
\sigma }\right) +\delta A_{\sigma }\left( -\varphi D_{\overline{z}}A_{\tau
}\right) +\delta A_{\overline{z}}\left( \varphi D_{\sigma }A_{\tau }\right)
\big\rangle ,
\end{equation}
where
\begin{equation}
d_{Vol}=d\sigma\wedge dz\wedge d\overline{z}.
\end{equation}

Now, we run the Dirac algorithm. The are three primary constraints given by%
\begin{equation}
P_{\tau }\approx 0,\text{ \ \ }\phi _{\sigma }=A_{\overline{z}}-\frac{8\pi }{%
i\varphi }P_{\sigma }\approx 0,\text{ \ \ }\phi _{\overline{z}}=P_{\overline{%
z}}+\frac{i\varphi }{8\pi }A_{\sigma }\approx 0.
\end{equation}%
The constraints $\phi _{\sigma }$ and $\phi _{\overline{z}}$ form a second
class pair and it is convenient to impose them strongly through a Dirac bracket before
we continue our analysis\footnote{Doing this at this level does not affect the final
outcome of the Dirac procedure but rather avoids extra and unnecessary computational effort.}. We change the Poisson brackets \eqref{PB CS} by
their corresponding Dirac brackets (DB) and, for $i=\sigma, \overline{z}$, both DB brackets boil down to 
\begin{equation}
\big\{ A_{\sigma}(\sigma ,z)_{%
\mathbf{1}},A_{\overline{z}}(\sigma^{\prime} ,z^{\prime})_{\mathbf{2}}\big\} ^{\ast }=\frac{4\pi }{i\varphi (z^{\prime})}C_{\mathbf{12}}\delta
_{\sigma \sigma^{\prime} }\delta _{zz^{\prime}}, \label{DB CS}
\end{equation}
while the Poisson bracket for $i=\tau$ remains unaltered. In what follows, we will drop the $*$ and continue referring to them simply as Poisson brackets in order to match common jargon. 

Using the remaining primary constraint, we construct the total Hamiltonian%
\begin{equation}
h_{T}=h_{C}+\int_{S^{1}\times 
\mathbb{C}
P^{1}}d_{Vol}\left\langle u_{\tau }P_{\tau }\right\rangle ,
\end{equation}%
where $u_{\tau }$ is an arbitrary Lagrange multiplier. 

The time preservation of the primary constraint $P_{\tau}\approx 0$ under the time evolution of $h_{T}$ leads to a secondary constraint
\begin{equation}
F\approx 0, \label{secondary F}
\end{equation}%
which is the first bulk eom written in \eqref{bulk eom}. In order to understand the geometric nature of this
constraint, its relation to the gauge symmetry of the theory and its role in the reduction process, it is convenient to invoke the symplectic approach before we continue.

Consider the symplectic form
associated to the Poisson brackets \eqref{DB CS}, which is given by a variant of the conventional CS symplectic form. It is given by
\begin{equation}
\hat{\Omega} =-\frac{i}{8\pi }\int\nolimits_{S^{1}\times 
\mathbb{C}
P^{1}}\omega \wedge \big\langle \hat{\delta }A\wedge \hat{\delta }%
A\big\rangle ,\text{ \ \ }A\in \mathcal{A}, \label{symplectic CS}
\end{equation}%
where $\hat{\delta }$ represents the exterior derivative in the symplectic manifold $\mathcal{A}$. Now, using the contraction%
\begin{equation}
\hat{\delta }A(X_{\eta })=-D\eta ,
\end{equation}%
where $X_{\eta }$ is the Hamiltonian vector field induced by the infinitesimal gauge
symmetry transformations \eqref{gauge transf} with $g=1+\eta$, $\eta\in \mathbf{g}=\Omega^{(0)}(S^{1}\times \mathbb{C}P^{1}, \mathfrak{f})$, we obtain%
\begin{equation}
-i_{X_{\eta }}\hat{\Omega} =\hat{\delta }H(\eta ),\text{ \ \ }H(\eta)\in\mathcal{C}^{\infty}(\mathcal{A}), \label{moment H}
\end{equation}%
where%
\begin{equation}
H(\eta )=\frac{i}{4\pi }\int\nolimits_{S^{1}\times 
\mathbb{C}
P^{1}}\omega \wedge \big\langle \eta F\big\rangle -\frac{i}{4\pi }%
\int\nolimits_{S^{1}\times 
\mathbb{C}
P^{1}}d\omega \wedge \big\langle \eta A\big\rangle \label{gauge ham} 
\end{equation}%
is the associated gauge Hamiltonian. Notice that \eqref{secondary F} constitutes the bulk contribution. A second contraction gives a centrally extended Poisson algebra%
\begin{equation}
\big\{ H(\eta ),H(\overline{\eta })\big\}=-H([\eta ,\overline{\eta }])-\frac{i}{4\pi }%
\int\nolimits_{S^{1}\times 
\mathbb{C}
P^{1}}d\omega \wedge \big\langle \eta d\overline{\eta }\big\rangle, \label{moment PB}
\end{equation}%
meaning that the gauge algebra must be centrally extended as well in order to have a morphism of Lie algebras. By equipping $\hat{\mathbf{g}}=\mathbf{g\oplus \mathbb{C}}$ with the cocycle\footnote{This is actually a collection of several contributions, one for each pole in $\mathfrak{p}$.}%
\begin{equation}
c(\eta ,\overline{\eta })=\frac{i}{4\pi }\int\nolimits_{
S^{1}\times \mathbb{C}P^{1}}d\omega\wedge \left\langle \eta d\overline{\eta }\right\rangle \label{CS cocycle}
\end{equation}%
and the bracket%
\begin{equation}
\left[ (\eta ,t),(\overline{\eta },s)\right] =([\eta ,\overline{\eta }%
],c(\eta ,\overline{\eta })) ,
\end{equation}%
we obtain a Lie algebra central extension of $\mathbf{g}$ and with the definition
\begin{equation}
H(\eta ,t)=H(\eta )+t, \label{H+t}
\end{equation} 
the mapping%
\begin{equation}
\begin{aligned}
\hat{\mathbf{g}} &\longrightarrow &\mathcal{C}^{\infty }(\mathcal{A}) \\
(\eta ,t) &\longmapsto &H(\eta ,t)
\end{aligned}%
\end{equation}
becomes a morphism of Lie algebras and we demand that $\hat{\mathbf{g}}$ generates the same gauge symmetry transformations \eqref{gauge transf}. This central extension is needed in order to accommodate the gauge symmetry in the correct way, as shown below in \eqref{center+gauge}.

Two comments are in order: i) the smeared constraint $H(\eta )$ has a
well-defined functional variation for any gauge parameter $\eta$, i.e.%
\begin{equation}
\delta H(\eta )=\frac{i}{4\pi }\int\nolimits_{S^{1}\times 
\mathbb{C}
P^{1}}\omega \wedge \big\langle \delta A\wedge D\eta \big\rangle, 
\end{equation}%
or more explicitly
\begin{equation}
\delta H(\eta )=-\frac{i}{4\pi }\int\nolimits_{S^{1}\times 
\mathbb{C}
P^{1}}d_{Vol}\big\langle \delta A_{\sigma }\{\varphi D_{\overline{z}}\eta
\}-\delta A_{_{\overline{z}}}\{\varphi D_{\sigma }\eta \}\big\rangle .
\end{equation}
Using the Poisson brackets \eqref{DB CS}, we write
the infinitesimal gauge symmetry transformations in Poisson form%
\begin{equation}
\big\{H(\eta ), A\big\} =-D\eta \label{Z}
\end{equation}%
and this means that $H(\eta )$ is identified as the gauge symmetry generator of the theory. However, ii) the
constraint algebra \eqref{moment PB} is first class only when 
$\eta |_{\mathfrak{p}}=0$, which is the second condition we found before in \eqref{gauge conditions} for the holomorphic CS theory action to be gauge
invariant and, as a consequence, true gauge symmetry transformations are generated only by the gauge parameters $\eta $ that vanish at the poles \eqref{poles}
of the twist function. Then, the first class smeared constraint of the theory denoted by
\begin{equation}
H_{0}({\eta}),\text{ \ \ }\eta|_{\mathfrak{p}}=0 \label{first class H}
\end{equation}
is precisely \eqref{secondary F} and not only the constraint is important but the pair $(F,\eta|_{\mathfrak{p}})$ is what matters. Considering now the gauge algebra
\begin{equation}
\mathbf{g}_{0}=\{\eta\subset \mathbf{g}|\text{\,}\eta|_{\mathfrak{p}}=0\}, \label{gauge-algebra}
\end{equation}
one realize that its corresponding gauge group $\mathcal{G}_{0}$ is normal. As a consequence, and in complete analogy to the situation considered in \cite{lambdaCS2} (see \cite{Aubin} for further details), after performing a SR and obtaining the space $\mathcal{A}_{0}$ of flat connections modulo gauge transformation generated by $\mathcal{G}_{0}$, there will be a residual gauge symmetry generated by $\mathcal{G}^{\prime}=\mathcal{G} / \mathcal{G}_{0}$ acting on it and the reduced space of the theory is actually the space $\mathcal{A}_{\text{red}}=\mathcal{A}_{0}/\mathcal{G}^{\prime}$. As we shall see, the space $\mathcal{A}_{0}$ naturally localizes at the set of poles \eqref{poles}, where the lambda deformed PCM starts to emerge. As an abuse of language, in the subsequent subsections we will use the label ``red'' in all quantities taking values in the space $\mathcal{A}_{0}$, the reason being a subtlety related to the residual gauge symmetry $\mathcal{G}^{\prime}$ and its role played in the lambda deformed PCM, so care must be taken. We comment on this in the paragraph above equation \eqref{XXX} below.

After this digression, we now continue with the Dirac procedure. Adding the secondary constraint \eqref{gauge ham} to the total Hamiltonian we construct the extended Hamiltonian
\begin{equation}
h_{E}=h_{T}+H(\overline{\eta}), \label{extended Ham}
\end{equation}
where the test function $\overline{\eta}$ plays the role of an arbitrary Lagrange multiplier. It is important to notice that $\overline{\eta}$ is not required to vanish on $\mathfrak{p}$ and that $h_{E}$ has a well-defined functional variation. 

The time preservation of the secondary constraint $H(\eta )$ under the time evolution of $%
h_{E}$ does not produce any further constraints but rather enforce the
condition $\eta |_{\mathfrak{p}}=0$ and, not surprisingly, only the first class constraint \eqref{first class H} is preserved in time. The primary constraint $P_{\tau }\approx 0$
is also preserved under the time evolution of $h_{E}$ and no tertiary constraints are
produced at this level. 

The only constraints of the theory are both first class%
\begin{equation}
P_{\tau }\approx 0,\text{ \ \ }H_{0}(\eta )\approx 0 \label{remaining 2}
\end{equation}%
and must be gauge fixed accordingly. However, we will only gauge fix the first one and
subsequently perform a SR with the second one and this is quite natural from the symplectic geometry point of view because of \eqref{secondary F} is not only a Hamiltonian constraint but also a piece of the moment map for the gauge symmetry as well. To see this, let us notice that \eqref{gauge ham} can be written in terms of a pairing\footnote{This pairing is assumed to be non-degenerated but problems might appear at the zeroes or the poles of the twist function, so we proceed formally.} between $\Omega ^{2}(S^{1}\times \mathbb{C}P^{1},%
\mathfrak{f})\oplus \Omega ^{1}(S^{1}\times \mathfrak{p},\mathfrak{f})\oplus \mathbb{C} $ and $\hat{\mathbf{g}}$ via
\begin{equation}
\big\langle(F,A,z),(\eta ,t)\big\rangle\longrightarrow \frac{i}{4\pi }%
\int\nolimits_{S^{1}\times \mathbb{C}P^{1}}\omega\wedge\big\langle  \eta F\big\rangle -\frac{i}{4\pi }%
\int\nolimits_{S^{1}\times \mathbb{C}P^{1}}d\omega\wedge\big\langle \eta A\big\rangle +zt. \label{pairing}
\end{equation}%
From this, we identify $\Omega ^{2}(S^{1}\times \mathbb{C}P^{1},%
\mathfrak{f})\oplus \Omega ^{1}(S^{1}\times \mathfrak{p},\mathfrak{f})\oplus \mathbb{C} $ as a subspace of $\hat{\mathbf{g}}^{\ast }$ and the mapping%
\begin{equation} 
\begin{split}
\mu\ :\quad &\mathcal{A} \longrightarrow \ \hat{\mathbf{g}}^{\ast } \\
&A \longmapsto \ (F,A|_{S^{1}\times \mathfrak{p}},1)
\end{split}
\end{equation}
is an equivariant moment map for the gauge group action because under \eqref{gauge transf}, we have that
\begin{equation}
H(\eta ,t)_{g}=H(Ad_{g^{-1}}(\eta,t)), \label{equivariance}
\end{equation}%
where%
\begin{equation}
Ad_{g}(\eta,t)=(Ad_{g}\eta, t_{g}),\text{ \ \ }t_{g}=t-\frac{i}{4\pi }\int\nolimits_{S^{1}\times \mathbb{C}P^{1}}d\omega\wedge \left\langle
\eta g^{-1}dg\right\rangle . \label{center+gauge}
\end{equation}

The first class constraint $P_{\tau }\approx 0$ can be gauged fixed via a
generic condition of the form%
\begin{equation}
A_{\tau }-\mathscr{L}_{\tau }(A_{\sigma},A_{\overline{z}})\approx 0. \label{fixing Ptau}
\end{equation}%
The only property we imposed on $%
\mathscr{L}_{\tau }(A)$ is that at the points $\mathfrak{p}$ the
boundary conditions \eqref{bdry CS} must be satisfied. This is a good gauge fixing condition
whose time preservation determines the Lagrange multiplier $u_{\tau }$ but as
it couples with the constraint $P_{\tau }$ in $h_{E}$, which is to be imposed strongly at the end anyway, its explicit form is not
relevant anymore. Furthermore, the PB \eqref{DB CS} is not modified by this gauge fixing and we are left only with the second constraint in \eqref{remaining 2}. 

We are now in the position to perform the SR and to show that the reduced CS theory corresponds to the lambda deformed PCM. We break the proof into three pieces, each one considering a relevant aspect of the reduced theory that is to be compared against the results gathered in \eqref{2}.

\subsubsection{Reduced Poisson structure: Maillet bracket}

The moment map for the $\mathcal{G}_{0}$-action of the normal gauge subgroup is given by the composition 
\begin{equation}
\begin{split}
&\mathcal{A}\ \: \overset{\mu }{\mathcal{\longrightarrow }}\qquad \ \ \ \hat{\mathbf{g}}%
^{\ast }\qquad \ \ \ \overset{p}{\mathcal{\longrightarrow }}\quad\ \hat{\mathbf{g}}_{0}^{\ast } \\
&A\ \longmapsto\ (F,A|_{S^{1}\times \mathfrak{p}},1)\ \longmapsto\ (F,1)
\end{split}
\end{equation}%
and $\mathcal{A}_{0}=(p\circ \mu )^{-1}(0,1)/\mathcal{G}_{0}$ is a symplectic reduced space. The $\mathcal{A}_{0}$ symplectic form of the theory is found by pulling-back \eqref{symplectic CS} to
the surface defined by the flatness condition $F=0$. Setting
\begin{equation}
A=-d\Psi \Psi ^{-1}, \label{flat A}
\end{equation}
we get that (c.f. \eqref{flat hol CS})%
\begin{equation}
\begin{aligned}
\hat{\Omega} _{\text{red}} &=-\frac{i}{8\pi }\int\nolimits_{S^{1}\times 
\mathbb{C}
P^{1}}d\omega \wedge \big\langle \Psi ^{-1}\hat{\delta }\Psi \wedge
d( \Psi ^{-1}\hat{\delta }\Psi ) \big\rangle  \\
&=\frac{i}{8\pi }\int\nolimits_{S^{1}\times 
\mathbb{C}
P^{1}}d\sigma \wedge d\omega \big\langle \hat{\delta }A_{\sigma }\wedge
D_{\sigma }^{-1}( \hat{\delta }A_{\sigma }) \big\rangle. 
\end{aligned}
\end{equation}%
Equivalently, from \eqref{useful} we find the important result
\begin{equation}
\hat{\Omega} _{\text{red}}=\frac{k}{4\pi }\int\nolimits_{S^{1}}d\sigma
\big\langle \hat{\delta }A_{\sigma }(z_{+})\wedge D_{\sigma
(+)}^{-1}( \hat{\delta }A_{\sigma }(z_{+})) -\hat{\delta }%
A_{\sigma }(z_{-})\wedge D_{\sigma (-)}^{-1}( \hat{\delta }%
A_{\sigma }(z_{-})) \big\rangle ,
\end{equation}%
where $D_{\sigma (\pm )}(\ast )=\partial _{\sigma }(\ast )+[A_{\sigma
}(z_{\pm }),\ast ]$ are covariant derivatives and $D_{\sigma (\pm )}^{-1}$ their formal inverses. At this point we have imposed the boundary condition \eqref{psi at infty} on $\Psi$ above, anticipating the validity of the result \eqref{A=L}. 
As a consequence of the SR, the symplectic form localizes at the poles of
the twist function where the reduced field theory phase space is now determined by the restricted CS field
\begin{equation}
A_{\sigma}(z_{\pm})=A_{\sigma}(z)|_{z=z_{\pm}}.
\end{equation} 

As mentioned before, there is a residual gauge symmetry action on the space $\mathcal{A}_{0}$
parameterized now by the coordinates $A_{\sigma }(z_{\pm })$. Indeed, using the contractions%
\begin{equation}
\hat{\delta }A_{\sigma }(z_{\pm })(X_{\eta })=-D_{\sigma (\pm )}\eta
(z_{\pm }),
\end{equation}%
where $\eta_{\pm}\in \mathbf{g}^{\prime}$, we find that%
\begin{equation}
-i_{X_{\eta }}\hat{\Omega} _{\text{red}}=\hat{\delta }H_{\text{red}}(\eta ), \label{1 con}
\end{equation}%
where%
\begin{equation}
H_{\text{red}}(\eta )=-\frac{1}{2}\sum\limits_{x\in \mathfrak{p}%
}res_{x}\omega \int\nolimits_{S^{1}}d\sigma \big\langle \eta A \big\rangle. \label{bdry Gauge}
\end{equation}
A second contraction in \eqref{1 con} gives their Poisson algebra%
\begin{equation}
\{H_{\text{red}}(\eta ),H_{\text{red}}(\overline{\eta} )\}=-\frac{1}{2}\sum\limits_{x\in \mathfrak{p}%
}res_{x}\omega \int\nolimits_{S^{1}}d\sigma \big\langle \eta D_{\sigma}%
\overline{\eta } \big\rangle,
\end{equation}
which is equivalent to two copies of mutually commuting KM
algebras of opposite levels,
\begin{equation}
\big\{A_{\sigma }(\sigma,z_{\pm} )_{\mathbf{1}},A_{\sigma }(\sigma^{\prime},z_{\pm} )_{\mathbf{2}}\big\}=\mp
\frac{2\pi }{k}\big( [C_{12},A_{\sigma }(\sigma^{\prime},z_{\pm} )_{\mathbf{2}%
}]\delta _{\sigma \sigma^{\prime} }+C_{12}\delta^{\prime} _{\sigma \sigma^{\prime} }\big). \label{KM from CS}
\end{equation}%
The expression \eqref{bdry Gauge} is the boundary contribution to the gauge Hamiltonian \eqref{gauge ham} and generates infinitesimal gauge symmetry transformations that can be written in Poisson form under \eqref{KM from CS}, 
\begin{equation}
\{H_{\text{red}}(\eta ),A_{\sigma}(z_{\pm})\}=-D_{\sigma(\pm)}\eta(z_{\pm}).
\end{equation}

Then, as a consequence of the SR procedure, in the reduced CS theory the $z$-dependent gauge field $A_{\sigma }(z)$
naturally interpolates between $A_{\sigma }(z_{+})$ and $A_{\sigma }(z_{-})$ so the obvious expression to be considered is inspired by the Lax connection \eqref{spatial-time Lax}%
\begin{equation}
A_{\sigma }(z)=\frac{k}{2\pi }f_{-}(z)A_{\sigma }(z_{+})-\frac{k}{2\pi }f_{+}(z)A_{\sigma }(z_{-}),\label{A(z)}
\end{equation}%
with the functions $f_{\pm}(z)$ defined as in \eqref{efes y ges}. As showed above around equation \eqref{MailletPCM}, the KM algebra structure induces the Maillet bracket \eqref{MailletPCM} on the component $
A_{\sigma }(z)$ and from \eqref{A(z)}, we obtain the identification
\begin{equation}
A_{\sigma}(z)=\mathscr{L}_{\sigma }(z),  
\end{equation}
justifying equation \eqref{A=L} for $\mu=\sigma$. The spectral parameter in \eqref{A(z)} now behaves as an auxiliary parameter, which is pretty much its usual interpretation in the classical theory.

Alternatively, if instead of performing the symplectic reduction, we choose to gauge fix the first class constraint $H_{0}(\eta)$ through the gauge fixing condition
\begin{equation}
A_{\overline{z}}\approx 0,
\end{equation}
the resulting Dirac bracket is, as shown in \cite{Vicedo-Holo}, the Maillet algebra bracket again. Hence, by fixing the gauge as right above or by performing the SR we get the same answer, the resulting reduced field theory being independent of the $A_{\overline{z}}$ component of the original CS gauge field. Also notice that in the SR procedure of the holomorphic CS theory, the component $A_{\overline{z}}$ behaves quite in the same way as the radius component $A_{r}$ of the gauge field in the SR of the double CS theory defined on the solid cylinder. 

\subsubsection{Reduced space equations of motion: Lax connection}

Let us identify the $\mathcal{A}_{\text{red}}$ symplectic leaves and the reduced eom on this space. 

Start with the symplectic leaves, which are determined by the action of the residual gauge algebra $\hat{\mathbf{g}}^{\prime}$ on $\mathcal{A}_{0}$. To find them, write \eqref{bdry Gauge} as a pairing between $\Omega ^{1}(S^{1}\times \mathfrak{p},\mathfrak{f})\oplus \mathbb{C} $ and $\hat{\mathbf{g}}^{\prime}$ of the form 
\begin{equation}
\big\langle(A,z),(\eta ,t)\big\rangle\longrightarrow -\frac{i}{4\pi }%
\int\nolimits_{S^{1}\times \mathbb{C}P^{1}}d\omega\wedge\big\langle \eta A\big\rangle +zt 
\end{equation}%
and identify $\Omega ^{1}(S^{1}\times \mathfrak{p},\mathfrak{f})\oplus \mathbb{C} $ as a subspace of $\hat{\mathbf{g}}^{\prime\ast }$. From the equivalence
\begin{equation}
\left\langle Ad_{g}^{\ast }(A,z),(\eta ,t)\right\rangle  =\left\langle
(A,z),Ad_{g^{-1}}(\eta ,t)\right\rangle
\end{equation}%
we have
\begin{equation}
Ad_{g}^{\ast }(A,z)=(Ad_{g}A-zdgg^{-1},z). \label{coad action}
\end{equation}
The case of interest is $z=1$, corresponding to the gauge transformations \eqref{gauge transf} restricted to $S^{1}\times\mathfrak{p}$. Then, the symplectic leaves in the reduced space are in one-to-one correspondence with the co-adjoint orbits \eqref{coad action}. A clearer picture appears by considering the transport matrix for the fields $A(z_{\pm})$, denoted generically by  
\begin{equation}
T(A|\sigma _{2},\sigma _{1})=P\exp \big[ -\int\nolimits_{\sigma
_{1}}^{\sigma _{2}}d\sigma A_{\sigma }(\sigma)\big] .
\end{equation}%
Under the co-adjoint action \eqref{coad action} with $z=1$, we have that 
\begin{equation}
T(Ad_{g}^{\ast }A|\sigma _{2},\sigma _{1})=g(\sigma _{2})T(A|\sigma
_{2},\sigma _{1})g(\sigma _{1})^{-1}
\end{equation}%
and the action on the monodromy matrix $m(A)=T(A|2\pi ,0)$ (at each point $z=z_{\pm}$) is given by 
\begin{equation}
m(Ad_{g}^{\ast }A)=g(0)m(A)g(0)^{-1},
\end{equation}%
showing that the co-adjoint orbits are in one-to-one correspondence with the orbits of the Lie group $%
F$ acting on itself by conjugation. Thus, the symplectic form \eqref{symplectic CS} induces a Poisson structure on $\mathcal{A}_{\text{red}}$, the symplectic leaves are then obtained by fixing the conjugacy classes of the monodromy matrices $m(A(z_{\pm}))$ along $S^{1}$. The $m(A(z_{\pm}))$ being related to finite-dimensional quantum groups \cite{quantum-group,Hidden-QG}. 

As $A_{\sigma}(z)$ is identified with the component $\mathscr{L}_{\sigma}(z)$, an important comment concerning the boundary residual gauge symmetry of the reduced theory is in order. Despite of the fact that \eqref{bdry Gauge} is interpreted as a gauge symmetry generator from the CS theory point of view, it is not a genuine gauge symmetry generator from the lambda deformed PCM perspective, as can be observed from \eqref{second class pair}, which states that no gauge symmetries are present in the theory and caution must be taken about its interpretation. This apparent enhancement of symmetry is also present in lambda models on semi-symmetric spaces \cite{lambdaCS2}, their true gauge symmetry being generated only by a subgroup of the residual CS theory gauge group and this could be understood as a being a consequence of embedding the lambda model phase space into a phase space of bigger dimension. Another issue, is that the gauge symmetry generated $\hat{\mathbf{g}}^{\prime}$ can not \cite{lambdaCS2} be continued outside the poles to act on  $A_{\sigma}(z)$ in the usual way as in \eqref{Z}. 

Now we consider the eom. In the reduced theory the gauge fixing condition \eqref{fixing Ptau} is now given by the strong expression
\begin{equation}
A_{\tau }=\mathscr{L}_{\tau }(A_{\sigma }), \label{XXX}
\end{equation}%
because of the independence of the reduced theory phase space on the coordinate $A_{%
\overline{z}}$. An explicit form that satisfies the boundary condition \eqref{bdry CS}, so far assumed to hold, is
clearly inspired by the Lax connection again. Thus, we take
\begin{equation}
A_{\tau }(z)=\frac{k}{2\pi }g_{-}(z)A_{\sigma }(z_{+})-\frac{k}{2\pi }g_{+}(z)A_{\sigma }(z_{-}),
\end{equation}%
with the functions $g_{\pm }(z)$ as defined in \eqref{efes y ges}, justifying equation \eqref{A=L} for $\mu=\tau$. 

The time evolution in the reduced theory is determined by the reduced Hamiltonian
\begin{equation}
h_{\text{red}}=-\frac{1}{4}\sum_{x\in \mathfrak{p}}res_{x}\omega\int\nolimits_{S^{1}}d\sigma\big\langle A%
_{\tau }A_{\sigma }\big\rangle,  \label{red Ham}
\end{equation}%
where we have taken $\overline{\eta }=0$ in \eqref{extended Ham} in order to separate it from
the contribution of the boundary gauge generator \eqref{bdry Gauge} to the extended Hamiltonian. This latter expression is the boundary contribution to
the canonical Hamiltonian \eqref{can Ham CS} and should be compared with the first equation in\footnote{The opposite sign when compared with \eqref{H and P as res} is not an issue as the equations \eqref{flow in lambda} and \eqref{flow in CS} are compatible
\begin{equation*}
\partial_{\tau}f=\{f,h_{\text{red}}\}=\{h_{C},f\}.
\end{equation*}
} \eqref{H and P as res}. By considering the
generator of translations along the $\sigma $ direction, given by
the second equation in \eqref{H and P as res}, we conclude that the pair of components $A_{\mu
}(z)$ of the reduced CS gauge field is actually a strongly flat $z$-dependent connection with respect to the KM algebra
structure of the reduced phase space and, as a consequence, we have that
\begin{equation}
A_{\tau }=-\partial _{\tau }\Psi \Psi ^{-1}. \label{flat Atau}
\end{equation}%

The $z$-dependent eom in the reduced space are given by the second expression in \eqref{bulk eom}, namely%
\begin{equation}
\partial _{\tau }A_{\sigma }-\partial _{\sigma }A_{\tau }+[A_{\tau
},A_{\sigma }]=0,\text{ \ \ }\partial _{\tau }A_{\overline{z}}-\partial_{\overline{z}}A_{\tau }+[A_{\tau },A_{\overline{z}}]=0.
\end{equation}%
The first equation giving the eom of the lambda deformed PCM \eqref{strong flatness}, while the second becomes an identity. Then, two of the components of the CS gauge field $B$, i.e. $(A_{\tau},A_{\sigma})$, behave as the lambda deformed PCM Lax connection for all intents and purposes.

Once we have understood the time evolution and the integrability of the eom in the reduced theory, we proceed to compute the corresponding action functional from where these quantities can be derived by canonical methods.

\subsubsection{Reduced action functional: lambda deformed PCM} 

In order to construct the reduced action functional having
\begin{equation}
\partial _{\tau }A_{\sigma }(z)-\partial _{\sigma }A_{\tau }(z)+[A_{\tau
}(z),A_{\sigma }(z)]=0
\end{equation}
as Euler-Lagrange\footnote{At least when restricted to set of poles, when it becomes \eqref{F eom}. Another signal of the localization in the reduced theory.} eom and \eqref{red Ham} as Hamiltonian function, we take 
\begin{equation}
B=-\hat{d}\Psi \Psi ^{-1}
\end{equation}%
into the action \eqref{Hol CS action} and this is because of equations \eqref{flat A} and \eqref{flat Atau}. We quickly find that%
\begin{equation}
S_{\text{red}}=-\frac{i}{8\pi }\int\nolimits_{\Sigma \times 
\mathbb{C}
P^{1}}\omega \wedge \chi (\Psi ).
\end{equation}%
This is not an standard WZ term because it involves an integral over the two-dimensional world-sheet manifold $\Sigma $
rather than on a three-dimensional manifold $\mathcal{M}$ with the property that $\Sigma =\partial 
\mathcal{M}$. To obtain an expression closer to the usual form, we denote by $\Psi ^{\prime }$ the extension of the wave
function into the five-dimensional manifold $\mathcal{M}\times \mathbb{C}P^{1}$ and write the result as an integral over $\mathcal{M}$. We get%
\begin{equation}
S_{\text{red}}=-\frac{i}{8\pi }\int\nolimits_{\mathcal{M}\times \mathbb{C}P^{1}}[ d\omega \wedge \chi (\Psi ^{\prime })-\omega \wedge \hat{d}%
^{\prime }\chi (\Psi ^{\prime })] ,
\end{equation}%
where%
\begin{equation}
\hat{d}^{\prime }=\hat{d}+dr\wedge \partial _{r}
\end{equation}%
is the extended exterior derivative with $r$ denoting the new coordinate. If
we further impose the condition
\begin{equation}
\int\nolimits_{\mathcal{M}\times \mathbb{C}P^{1}}\omega \wedge \hat{d}%
^{\prime }\chi (\Psi ^{\prime })=0 ,\label{important}
\end{equation}  
e.g. a condition satisfied if the WZ three-form is closed under $\hat{d}%
^{\prime }$, the reduced action functional takes the form
\begin{equation}
S_{\text{red}} =-\frac{k}{4\pi }\int\nolimits_{\mathcal{M}}[\chi (\Psi ^{\prime
}(z_{+}))-\chi (\Psi ^{\prime }(z_{-}))], \label{eff from CS}
\end{equation}
which is equal to the action found before in \eqref{lambda model action}. This same action was found in \cite{lambdaCS2} starting from the double CS theory. Notice that under the validity of \eqref{important}, the first condition in \eqref{gauge conditions} is satisfied giving $N=0$. We will assume that equation \eqref{important} holds, although we will comment about its validity in the final remarks section below.

A more suggestive expression follows from the relations \eqref{key relations} found above. Indeed, in terms of the field
variables 
\begin{equation}
\mathcal{F}=\Psi(z_{+})\Psi(z_{-})^{-1}\text{ \ \ and \ \ } \mathcal{F}^{\prime}=\Psi^{\prime}(z_{+})\Psi^{\prime}(z_{-})^{-1} \label{fields at poles}
\end{equation}  
defined on $\Sigma$ and $\mathcal{M}$, respectively, the reduced action becomes \cite{lambdaCS2}
\begin{equation}
S_{\text{red}}=\frac{k}{2\pi }\int\nolimits_{\Sigma}d^{2}\sigma \left\langle \mathcal{F}^{-1}\partial _{+}\mathcal{F}\mathscr{L}%
_{-}(z_{-})-\mathcal{F}^{-1}\partial _{-}\mathcal{F}\mathscr{L}_{+}(z_{-})\right%
\rangle -\frac{k}{4\pi }\int\nolimits_{\mathcal{M}}\chi (\mathcal{F^{\prime}}) \label{eff action}
\end{equation}%
or, alternatively,
\begin{equation}
S_{\text{red}}=\frac{k}{2\pi }\int\nolimits_{\Sigma }d^{2}\sigma \left\langle \partial _{+}\mathcal{F}\mathcal{F}^{-1}\mathscr{L}%
_{-}(z_{+})-\partial _{-}\mathcal{F}\mathcal{F}^{-1}\mathscr{L}_{+}(z_{+})\right%
\rangle -\frac{k}{4\pi }\int\nolimits_{\mathcal{M}}\chi (\mathcal{F^{\prime}}). \label{eff action 2}
\end{equation}%
Both expressions being equivalent under the symmetry \eqref{rare symmetry}. This symmetry which manifests trivially in the CS theory formulation was first discovered in \cite{Sfetsos-Thirring} by working directly on the effective action \eqref{final} written below, has important implications for the renormalization group structure of the theory, see \cite{Sfetsos-Thirring}.

Expressions \eqref{eff action} and \eqref{eff action 2} can be written in terms of residues over the poles \eqref{poles}. In order to do this, we introduce a $z$-dependent extension of the Lagrangian fields \eqref{fields at poles} defined by
\begin{equation}
\mathscr{F}(z)=\Psi(-z)\Psi(z)^{-1}\text{ \ \ and \ \ } \mathscr{F}^{\prime}(z)=\Psi^{\prime}(-z)\Psi^{\prime}(z)^{-1}
\end{equation}  
and, as a consequence, we have that
\begin{equation}
\mathcal{F}=\mathscr{F}(z_{\pm})^{\mp 1}.
\end{equation} 
The expressions \eqref{eff action 2} and \eqref{eff action}, respectively, now take the compact form
\begin{equation}
S_{\text{red}}=\frac{1}{4}res_{z_{\pm }}\omega \Big( \int_{\Sigma}\left\langle \mathscr{F}^{-1}d\mathscr{F\wedge L}%
\right\rangle +\int_{\mathcal{M}}\mathscr{\chi (F}^{\prime }\mathcal{)}\Big), 
\end{equation}%
where
\begin{equation}
\mathscr{L}(z)=-d\Psi(z)\Psi(z)^{-1}
\end{equation}
is the strongly flat Lax connection of the theory and from this we obtain the final form for the reduced action functional
\begin{equation}
S_{\text{red}}\equiv \underset{x\in \mathfrak{p}}{\sum }S_{x}=\frac{1}{8}\underset{x\in \mathfrak{p}}{\sum }res_{x}\omega 
\Big( \int_{\Sigma}\left\langle \mathscr{F}^{-1}d\mathscr{F\wedge L}%
\right\rangle +\int_{\mathcal{M}}\mathscr{\chi (F}^{\prime }\mathcal{)}\Big).
\end{equation}
A similar expression was found in \cite{Vicedo-Unif} to be valid as well for a wide range of other integrable deformations of string sigma models, after fixing a particular form of the CS gauge field (roughly) enforcing the condition $A_{\overline{z}}=0$ and imposing the so-called archipelago conditions on an analogue to the field $\mathscr{F}$ defined above. Here, the CS theory naturally localize at $\mathfrak{p}$, as a consequence of the SR.

Finally, an equivalent expression for the reduced action, say \eqref{eff action}, is given by \cite{lambdaCS2} 
\begin{equation}
S_{\text{red}}=\frac{k}{2\pi }\int\nolimits_{\Sigma 
}d^{2}\sigma \left\langle \mathcal{F}^{-1}\partial _{+}\mathcal{F}A_{-}-%
\mathcal{F}^{-1}\partial _{-}\mathcal{F}\Omega ^{T}A_{+}\right\rangle -\frac{%
k}{4\pi }\int\nolimits_{\mathcal{M}
}\chi (\mathcal{F^{\prime}}),
\end{equation}%
where $A_{\pm}$ depend on $\mathcal{F}$ via \eqref{A in terms of F}. After some algebraic manipulations, we obtain a more familiar form%
\begin{equation}
S_{\text{red}}=-\frac{k}{2\pi }\int\nolimits_{\Sigma
}d^{2}\sigma \left\langle \mathcal{F}^{-1}\partial _{+}\mathcal{F}(G+B)%
\mathcal{F}^{-1}\partial _{-}\mathcal{F}\right\rangle, \label{final}
\end{equation}%
where
\begin{equation}
\begin{aligned}
G &=\frac{1}{(\Omega -D)}(\Omega \Omega ^{T}-1)\frac{1}{(\Omega ^{T}-D^{T})}, \\
B &=B_{0}+\frac{1}{(\Omega -D)}(D\Omega ^{T}-\Omega D^{T})\frac{1}{(\Omega
^{T}-D^{T})},
\end{aligned}
\end{equation}%
with $B_{0}$ denoting the WZ term contribution. The ``kernels'' $G$ and $B$ are responsible for the deformation of the metric and anti-symmetric background fields. The action \eqref{final} is the effective action we obtain from \eqref{deformed-PCM} by integrating out the gauge fields $A_{\pm}$ through their equations of motion. Then,
\begin{equation}
S_{\text{eff}}=S_{\text{red}}=\underset{x\in \mathfrak{p}}{\sum }S_{x},
\end{equation}
showing that the theory localize at the poles of the twist function under the SR of its parent holomorphic CS theory. The symplectic reduced holomorphic CS theory being identified with the lambda deformed PCM.

\section{Final remarks}\label{4}

We have shown how the symplectic reduction applied to a particular holomorphic Chern-Simons theory works as a localization mechanism in the phase space of the theory. The physical data of the reduced theory associated to the points where the theory localize is sufficient to reconstruct the lambda deformation of the Principal Chiral Model and all of its known integrability properties. From the lessons of \cite{lambdaCS2}, this opens the possibility of considering lambda models on (semi)-symmetric spaces, like the $AdS_{5}\times S^{5}$ superstring lambda model, in a direct way and this is because of an analogue of the expression \eqref{A(z)} in that case is already known to induced the classical exchange (Maillet's) algebra for the theory. Other cases of lambda models from the point of view of the holomorphic CS theory will be considered in a companion work.  

Let us comment on the relation between approaches (I) and (II) mentioned in the introduction. By introducing a three-dimensional manifold $\mathcal{M}$ with the property that $\partial \mathcal{M}=\Sigma$ and by extending all the quantities \eqref{quantities} into this manifold (primed variables), the action \eqref{Hol CS action} becomes
\begin{equation}
S_{CS}=\frac{i}{8\pi }\int\nolimits_{\mathcal{M} \times \mathbb{C}P^{1}}d\omega \wedge CS(B^{\prime})-\frac{i}{8\pi }\int\nolimits_{\mathcal{M} \times \mathbb{C}P^{1}} \omega \wedge \big\langle F^{2}_{B^{\prime}} \big\rangle, \label{Hol CS action 2}
\end{equation}  
where $F_{B^{\prime}}$ is the field strength of the gauge field $B^{\prime}$. Under gauge symmetry transformations, we find the variation
\begin{equation}
\delta S_{CS}=\frac{i}{8\pi }\int\nolimits_{\mathcal{M}\times 
\mathbb{C}
P^{1}}d\omega \wedge \chi (g^{\prime})+\frac{i}{8\pi }\int\nolimits_{\Sigma \times \mathbb{C}P^{1}}d\omega \wedge \big\langle g^{-1}\hat{d}g\wedge B\big\rangle. \label{deltaCS}
\end{equation}%
Notice that this variation is completely localized at the set of poles $\mathfrak{p}$. \\
(I) \textit{``Holography''}. When $\mathcal{M}=\mathbb{R}\times D$ is the solid disc and $B^{\prime}$ is chosen to vanish at $z=\infty$, the first contribution to \eqref{Hol CS action 2} is precisely the double CS theory action functional of \cite{lambdaCS,lambdaCS2}. In this case, the relation between the physical information contained in the interior of the disc and its geometric boundary is what matters. For instance, gauge invariance in \eqref{deltaCS} leads to two conditions \cite{lambdaCS2}
\begin{equation}
\frac{k}{4\pi }\int\nolimits_{\mathcal{M}}[\chi (g^{\prime}(z_{+}))-\chi
(g^{\prime}(z_{-}))]=2\pi N,\text{ \ \ }g|_{\partial \mathcal{M}}=Id.
\end{equation}%
The first condition imposes a quantization condition while the second one imposes a boundary condition on the gauge group elements at the boundary $S^{1}$ of the disc. This interpretation is quite close to the known holographic results of \cite{Seiberg,zoo,Aubin}, where the SR reduction eliminates the dof in the interior of the disc but retains only those belonging to its geometric boundary. This standard interpretation is what initially inspired the approach (I) and the works \cite{lambdaCS,lambdaCS2}. \\
(II) \textit{``Localization''}. Because of the action \eqref{Hol CS action} is inherently defined on $\Sigma$ and no geometric boundary terms are present, what matters now is to discriminate the physical information associated to the poles from the rest, i.e. the ``bulk''. For instance, both contributions in \eqref{deltaCS} vanish simply by imposing the condition
\begin{equation}
g|_{\mathfrak{p}}=Id
\end{equation} 
on the gauge group elements and the first condition in \eqref{gauge conditions} is then unnecessary, in consistency with \eqref{gauge-algebra}. Let us notice as well that the equation \eqref{important} must hold because of the action \eqref{Hol CS action 2} becomes \eqref{eff from CS} when restricted to the set of flat gauge fields. \\
Fortunately, the results of \cite{Aubin} (dealing purely with surfaces with boundaries) still apply with minor modifications in this scenario giving the same reduced theory. It is the interplay among the objects $\mathcal{M}$, $\Sigma$, $\partial$ and $\hat{d}^{\prime}$ that allows to easily relate approaches (I) and (II) and to understand why they give the same classical reduced field theory. Both approaches match because of the action \eqref{Hol CS action 2} is to be restricted to the set of flat gauge fields, where the second term is absent.

Finally, as a first quantum test to show that the equivalence between the holomorphic CS theory and the lambda models indeed goes beyond the classical regime, it would be interesting to recover the lambda model dilaton term contribution to the action \eqref{final}, directly from the four-dimensional CS gauge theory \eqref{Hol CS action}. We expect to consider this issue elsewhere.

\section*{Acknowledgements}

The work of DMS is supported by the S\~ao Paulo Research Foundation (FAPESP) under the research grant 2017/25361-7.

\end{document}